%% file: main.tex
\documentclass{article}
\usepackage{spconf,amsmath,graphicx,hyperref}
\usepackage{booktabs,multirow,amssymb}
\usepackage{subfig}
\usepackage{microtype}
\usepackage{etoolbox}

\apptocmd{\thebibliography}{%
  \setlength{\itemsep}{0pt}%
  \setlength{\parsep}{0pt}%
  \setlength{\topsep}{0pt}%
  \setlength{\partopsep}{0pt}%
}{}{}

\title{CORE-DSE: Coarse Extraction and Refinement with Time-Frequency Gated Cross-Attention for
  Directional Speech Extraction\vspace{-4.5pt}}

\name{\shortstack{Yifei Yang$^{1,2,\dagger}$, Wentao Hua$^{1,2,\dagger}$, Jizhe Lu$^{1,2}$, Leyan Yang$^{1,2}$,\\
  Jun Gao$^{1,2}$, Yuxiang Hu$^{2}$, Changbao Zhu$^{2}$, Jing Lu$^{1,2,\star}$}\vspace{-4.5pt}}
\address{$^{1}$Key Laboratory of Modern Acoustics, Nanjing University, Nanjing 210093, China\\
  $^{2}$NJU-Horizon Intelligent Audio Lab, Horizon Robotics, Beijing 100094, China\\
  $^{\star}$\small\texttt{lujing@nju.edu.cn}\vspace{-1.75em}}

\begin{document}
\ninept

\maketitle
\begingroup
\renewcommand{\thefootnote}{\fnsymbol{footnote}}
\footnotetext[2]{\fontsize{9}{10}\selectfont Yifei Yang and Wentao Hua contributed equally to this work.}
\endgroup

\begin{abstract}
  Direction of arrival (DOA) is a natural cue for multichannel directional speech extraction (DSE),
  but its few global geometric variables are poorly matched to dense time-frequency acoustic
  representations, limiting the performance of existing DSE methods. We propose CORE-DSE, a coarse
  extraction and refinement framework that addresses this representation-granularity mismatch.
  Specifically, the Coarse Extractor (CE) combines Time-Frequency Gated Cross-Attention (TF-GCA),
  which models the relevance of the target direction to individual time-frequency units, with a Local Gated Mixer
  (LGM) to produce direction-aware spectral and spatial representations. A Fine Enhancer (FE) then
  suppresses residual interference and restores target speech details. Experiments show that CORE-DSE
  consistently outperforms physics-informed and generic DOA conditioning baselines with only marginal computational overhead. Effective directional fusion yields greater gains than
  increasing model capacity, and CE consistently improves performance across different enhancement
  backbones.
\end{abstract}


\section{Introduction}

Target speech extraction (TSE) isolates a target speaker from noisy, reverberant mixtures with
competing speakers using auxiliary cues such as enrollment speech, visual information, or source
location~\cite{zmolikova2023overview}. For microphone arrays,
direction of arrival (DOA), obtained through acoustic localization~\cite{wang2023fnssl} or visual
detection~\cite{jiang2023lip}, provides a natural spatial cue for directional speech extraction
(DSE), eliminating the need for enrollment speech when the target direction is
known~\cite{zhang2025doa}. This makes DSE attractive for video conferencing, human–machine
interaction, and robot audition~\cite{wang2026position,wang2022hybrid}.

Enrollment-based TSE has progressed from fixed global speaker
embeddings~\cite{wang2019voicefilter,xu2020spex} to multi-level, local, and contextual
representations, with cross-attention enabling fine-grained target-mixture
interactions~\cite{zhang2025multilevel,zeng2025useftse,yang2024coarse}. However, transferring this paradigm to DSE is challenging because enrollment-based representations and DOA cues differ substantially in structure. Speaker embeddings are learned from rich
spectro-temporal content in speech, while enrollment speech can further provide features at the frame,
spectral, or token level. In contrast, DOA contains only one or two geometric variables and remains a
deterministic basis expansion even after Fourier~\cite{alcalapadilla2025location} or spherical
harmonic~\cite{choi2026soundcompass} encoding. This results in a representation-granularity mismatch
between low-dimensional global directional priors and dense time-frequency (TF) acoustic
representations, hindering direct adaptation of enrollment-oriented interaction mechanisms.

Existing DSE methods generally address this mismatch in two ways. Physics-informed methods use array
geometry and acoustic propagation models to transform DOA into handcrafted matching features aligned
with TF observations, such as angle features~\cite{chen2018location} and beamforming
responses~\cite{elminshawi2023beamformer}. They provide explicit local matching, but rely on
accurate physical assumptions and steering information, and may suffer from mismatch under strong
reverberation or near-field conditions~\cite{luo2025steering}. Another class of methods directly
injects encoded DOA into the network through generic conditioning mechanisms such as feature
concatenation~\cite{alcalapadilla2025location}, hidden-state initialization~\cite{tesch2024spatial},
or feature-wise modulation~\cite{perez2018film}. Although flexible, they usually treat DOA as a
global condition without explicit interaction with individual TF components, leaving directional
awareness to be learned implicitly by subsequent networks. Thus, physics-informed features offer
fine-grained but predefined matching, whereas generic neural conditioning offers flexibility at
coarser granularity.

Based on these observations, we argue that effective DSE should better capture the fine-grained
relationships between global directional cues and local acoustic representations, rather than relying
solely on sophisticated DOA encoding or stronger enhancement backbones. To this end, we propose
Time-Frequency Gated Cross-Attention (TF-GCA), tailored to low-dimensional global directional
conditioning, to model the relevance of DOA to individual TF units in latent space and
mitigate the representation-granularity mismatch. We further introduce \textbf{CO}arse extraction and
\textbf{RE}finement (CORE), a framework decomposing DSE into direction-aware coarse extraction and
target-speech refinement. The Coarse Extractor (CE) integrates DOA through TF-GCA and a Local Gated
Mixer (LGM) to produce direction-guided spectral and spatial embeddings. The Fine Enhancer (FE)
suppresses residual interference and restores target-speech details using a speech enhancement
network. Experiments show that CORE-DSE outperforms multiple DOA conditioning strategies with marginal
computational overhead, while CE consistently improves performance across different FE backbones, demonstrating the effectiveness and generality of our framework.

\section{Proposed Method}

\input{Figures/fig1_layout}

\subsection{Problem Formulation}

Consider an $M$-channel microphone array observing a mixture of a target speaker, $J$ interfering
speakers, and noise. The signal at the $m$-th microphone is
\begin{equation}
  x_m[n]
  =
  s_m[n]
  +
  \sum_{j=1}^{J} u_{j,m}[n]
  +
  v_m[n],
\end{equation}
where $s_m[n]$ is the target speech, and $u_{j,m}[n]$ and $v_m[n]$ are the
$j$-th interfering speech and noise, respectively.

Applying the short-time Fourier transform (STFT), we obtain
$\mathbf{X}_{f,t}=[X_{0,f,t},\ldots,X_{M-1,f,t}]^\top\in\mathbb{C}^{M}$ at TF bin $(f,t)$.
Given the target DOA $\mathbf d=(\theta,\phi)$, with $\theta$ and $\phi$ denoting azimuth and
elevation, DSE estimates the target speech at the reference microphone as
\begin{equation}
  \widehat{\mathbf S}_0
  =
  \mathcal{F}_{\Theta}(\mathbf{X},\mathbf{d}),
\end{equation}
where $\mathbf X\in\mathbb C^{M\times F\times T}$ collects
$\{\mathbf X_{f,t}\}$ over all TF bins.

\subsection{CORE Framework}

As shown in Fig.~\ref{fig:core_dse_overview}, CORE-DSE decomposes DSE into direction-aware coarse
extraction and fine-grained speech enhancement. CE uses the target DOA to produce direction-constrained
latent acoustic representations, while FE further suppresses residual interference and restores
target-speech details. Both stages are jointly optimized under target-speech supervision.

Figure~\ref{fig:coarse_extractor} details the CE architecture. Given the multichannel STFT
observation $\mathbf X$, CE projects the concatenated real and imaginary components into the
spectral feature:
\begin{equation}
  \mathbf A^{\mathrm{spec}}
  =
  \mathcal P_{\mathrm{spec}}
  \left(
  [
  \operatorname{Re}(\mathbf X);
  \operatorname{Im}(\mathbf X)
  ]_{\mathrm c}
  \right)
  \in
  \mathbb R^{C_s\times F\times T},
  \label{eq:spectral_feature}
\end{equation}
where $[\cdot;\cdot]_{\mathrm c}$ denotes channel-wise concatenation and
$\mathcal P_{\mathrm{spec}}$ is the spectral feature projection.

For the $P$ microphone pairs $\{(m_p,n_p)\}_{p=1}^{P}$, the inter-channel phase differences (IPDs)
are stacked as
\begin{equation}
  \boldsymbol{\Delta\Psi}
  =
  \left[
  \angle X_{m_p,f,t}
  -
  \angle X_{n_p,f,t}
  \right]_{p=1}^{P}
  \in
  \mathbb R^{P\times F\times T}.
  \label{eq:ipd}
\end{equation}
Their sine-cosine representation is projected into the spatial feature:
\begin{equation}
  \mathbf A^{\mathrm{spat}}
  =
  \mathcal P_{\mathrm{spat}}
  \left(
  [
  \sin\boldsymbol{\Delta\Psi};
  \cos\boldsymbol{\Delta\Psi}
  ]_{\mathrm c}
  \right)
  \in
  \mathbb R^{C_p\times F\times T},
  \label{eq:spatial_feature}
\end{equation}
where $\mathcal P_{\mathrm{spat}}$ is the spatial feature projection. Mapping the wrapped phase onto
the unit circle avoids discontinuities at $-\pi$ and $\pi$~\cite{wang2019spectral}.

\input{Figures/fig2_layout}

The Condition Encoder in Fig.~\ref{fig:core_dse_overview} maps the target direction to a directional
feature using spherical harmonic (SH) encoding up to order $L$. Let
$\mathbf y_L(\theta,\phi)\in\mathbb C^{(L+1)^2}$ collect $Y_\ell^q(\theta,\phi)$ for
$0\leq\ell\leq L$ and $-\ell\leq q\leq\ell$ in a fixed order. The directional feature is obtained by
concatenating its real and imaginary parts:
\begin{equation}
  \mathbf A^{\mathrm{dir}}
  =
  \left[
  \operatorname{Re}(\mathbf y_L(\theta,\phi));
  \operatorname{Im}(\mathbf y_L(\theta,\phi))
  \right]_{\mathrm c}
  \in
  \mathbb R^{C_d},
  \label{eq:direction_feature}
\end{equation}
where $C_d=2(L+1)^2$ and $Y_\ell^q(\theta,\phi)$ is the complex SH function of degree
$\ell$ and order $q$. SHs provide a continuous, structured, and physically consistent representation
of three-dimensional directionality, with different orders capturing different angular
scales~\cite{choi2026soundcompass}.

\input{Figures/fig3_latout}

CE consists of separate spectral and spatial branches. For branch
$b\in\{\mathrm{spec},\mathrm{spat}\}$, TF-GCA matches the directional feature with individual TF
units, while LGM models their local context:
\begin{equation}
  \begin{gathered}
    \mathbf Z^{b}
    =
    \operatorname{LGM}_{b}
    \left(
    \operatorname{TF\mbox{-}GCA}_{b}
    (\mathbf A^{b},\mathbf A^{\mathrm{dir}})
    \right),\\
    \mathbf Z^{\mathrm{CE}}
    =
    [\mathbf Z^{\mathrm{spec}};
    \mathbf Z^{\mathrm{spat}}]_{\mathrm c}.
  \end{gathered}
  \label{eq:coarse_extraction}
\end{equation}
The fused feature $\mathbf Z^{\mathrm{CE}}$ retains the TF resolution and serves as a
direction-modulated coarse representation from which FE estimates the target speech at the reference
microphone:
\begin{equation}
  \widehat{\mathbf S}_0
  =
  \mathcal F_{\mathrm{FE}}
  (\mathbf Z^{\mathrm{CE}}).
  \label{eq:fine_enhancement}
\end{equation}
FE can use different speech enhancement networks and serves to suppress residual interference and restore target-speech details.

\subsection{Time-Frequency Gated Cross-Attention}

As shown in Fig.~\ref{fig:tf_gca_lgm}(a), TF-GCA models the fine-grained relevance between the
target direction and individual TF units. Standard softmax
cross-attention cannot yield the desired bin-wise modulation: a single
directional key makes every softmax weight $1$, whereas a directional query over all TF keys
produces one aggregated output. TF-GCA instead uses the directional feature as a global query and TF
units as local keys and values, with an independent sigmoid gate for each bin.

For $b\in\{\mathrm{spec},\mathrm{spat}\}$, the separately normalized
directional feature $\mathbf A^{\mathrm{dir}}$ and acoustic feature
$\mathbf A^b$ are projected into $H$ query heads and $H$ key/value
heads, respectively, using branch-specific projections. At $(f,t)$, the $h$-th head computes
\begin{equation}
  \begin{aligned}
    g_{h,f,t}^{b}
    &=
    \sigma\left(
    \frac{
    (\mathbf q_{h}^{b})^\top
    \mathbf k_{h,f,t}^{b}
    }{
    \sqrt{d_h}
    }
    \right)
    \in(0,1),\\
    \mathbf u_{h,f,t}^{b}
    &=
    g_{h,f,t}^{b}\mathbf v_{h,f,t}^{b},
  \end{aligned}
  \label{eq:tfgca_gate}
\end{equation}
where $\mathbf q_h^b$, $\mathbf k_{h,f,t}^b$, $\mathbf v_{h,f,t}^b$, and $\mathbf u_{h,f,t}^b$ are
in $\mathbb R^{d_h}$, and $\sigma(\cdot)$ is the sigmoid function. The gated values are merged,
projected, and residually added to the input:
\begin{equation}
  \widetilde{\mathbf A}^{b}
  =
  \mathbf A^{b}
  +
  \operatorname{Dropout}
  \left(
  \mathcal P_{O}^{b}
  \left(
  \mathbf U^{b}
  \right)
  \right),
  \label{eq:tfgca_output}
\end{equation}
where $\mathbf U^b=[\mathbf u_{1}^{b};\ldots;\mathbf u_{H}^{b}]_{\mathrm c}
\in\mathbb R^{Hd_h\times F\times T}$ collects the gated values from all heads, with
$\mathbf u_h^b=\{\mathbf u_{h,f,t}^b\}_{f,t}$ denoting the gated-value map of the $h$-th head.

In the spatial branch, TF-GCA learns per-bin DOA--IPD compatibility without explicitly constructing
the steering or phase templates used in physics-informed methods. The gates are computed independently and are not constrained to sum to $1$, allowing multiple
target-relevant TF units to be updated without cross-bin competition or aggregation. Multiple heads
learn complementary matching patterns in different feature subspaces, while the acoustic values and
residual connection preserve the TF grid and original input.

\subsection{Local Gated Mixer}

TF-GCA performs bin-wise directional matching without modeling acoustic continuity across adjacent
frames, so evidence from individual TF units may be less reliable under noise and reverberation. As
shown in Fig.~\ref{fig:tf_gca_lgm}(b), LGM follows TF-GCA to capture local temporal context and
perform gated mixing on the direction-modulated features.

For branch $b$, the normalized TF-GCA output $\widetilde{\mathbf A}^{b}$ is projected by a
$1\times1$ convolution and split along the channel dimension into content and gate features:
\begin{equation}
  \left[
  \mathbf C^{b};
  \mathbf G^{b}
  \right]_{\mathrm c}
  =
  \mathcal P_{\mathrm{in}}^{b}
  \left(
  \operatorname{LN}
  (\widetilde{\mathbf A}^{b})
  \right),
  \label{eq:lgm_split}
\end{equation}
where $\mathbf C^b,\mathbf G^b\in\mathbb R^{C_b'\times F\times T}$, $C_b'=rC_b$, and $r$ is the
channel expansion factor. The content branch applies a grouped one-dimensional temporal convolution
(T-GConv1D) to model local temporal context, while SiLU produces adaptive modulation features from
$\mathbf G^b$. Their element-wise product is projected and residually added to the input:
\begin{equation}
  \mathbf Z^{b}
  =
  \widetilde{\mathbf A}^{b}
  +
  \mathcal P_{\mathrm{out}}^{b}
  \left(
  \operatorname{Dropout}
  \left[
  \operatorname{TConv}
  (\mathbf C^{b})
  \odot
  \operatorname{SiLU}(\mathbf G^{b})
  \right]
  \right).
  \label{eq:lgm_output}
\end{equation}

T-GConv1D applies a $1\times K_t$ kernel along time at each frequency bin. Compared with
SwiGLU~\cite{shazeer2020glu} using only pointwise projections, LGM adds local temporal modeling at
low computational cost, complementing the bin-wise directional matching of TF-GCA in CE.

\section{Experiments}

\subsection{Experimental Setup}

We simulated two spatialized LibriSpeech~\cite{panayotov2015librispeech} datasets at
16~kHz using a six-microphone uniform circular array with a 3.6-cm radius and room impulse responses (RIRs) generated with
\textit{pyroomacoustics}~\cite{scheibler2018pyroomacoustics}. Noise was drawn from the DNS3
corpus~\cite{reddy2021dns3}. All mixtures were 10~s long; the \textbf{Easy} and \textbf{Hard}
datasets contained 60k/1k/1k and 100k/1k/1k training/validation/test samples, respectively.

For \textbf{Easy}, one target and one interfering speaker were placed on the array plane at
0.5--3.0~m with at least $30^\circ$ azimuth separation, in rooms of $3$--$10 \times 3$--$10
\times 2.5$--$3.0$~m with $T_{60}$ of 0.1--0.4~s. Point-source noise and the interferer were mixed
at SNRs of $[-10,5]$~dB and SIRs of $[-5,5]$~dB, respectively.

For \textbf{Hard}, 2--5 speakers were placed over the full sphere at 0.5--5.0~m with at least
$5^\circ$ pairwise angular separation. Small, medium, and large rooms were sampled equally with
$T_{60}$ spanning 0.1--1.0~s. Point-source and diffuse noise types were sampled with probabilities 0.9
and 0.1, with SNRs of $[-15,5]$~dB and SIRs of $[-10,10]$~dB.

Input mixtures were processed using a 512-point STFT with a 32~ms Hann window and a 16~ms hop. CE
used $C_s=C_p=64$, $L=5$, and $H=4$, while LGM
used $r=1$, $K_t=5$, and 16 groups in T-GConv1D. All models were trained with the hybrid
loss~\cite{rong2024gtcrn}:
\begin{equation}
  \mathcal{L}
  =0.3(\mathcal{L}_{\mathrm{R}}+\mathcal{L}_{\mathrm{I}})
  +0.7\mathcal{L}_{\mathrm{Mag}}
  +0.01\mathcal{L}_{\mathrm{SI\text{-}SDR}}.
  \label{eq:training_loss}
\end{equation}
Optimization used Adam~\cite{kingma2015adam} for 250 epochs with 10k samples per epoch and a global
batch size of 8. The learning rate was linearly warmed to $10^{-3}$ over the first 25 epochs and
cosine-decayed to $10^{-6}$.

\subsection{Comparison of DOA Conditioning Strategies}

\input{Tables/table1}

Table~\ref{tab:doa_conditioning} compares DOA conditioning strategies (IDs~1--8) on the
\textbf{Easy} dataset using FT-JNF~\cite{tesch2022ftjnf} as the common FE. IDs~1--2 concatenate
physics-informed features with the original input. For a fair comparison, IDs~4--7 share the
dual-branch structure in Fig.~\ref{fig:coarse_extractor}; IDs~4--6 differ only in DOA fusion,
whereas ID~7 increases model capacity of FE. Cross-attention (ID 6) repeats the DOA feature across TF bins and uses a fourfold FFN expansion. ID~8 retains FT-JNF as the FE and replaces CE with the SoundCompass Fusion Module~\cite{choi2026soundcompass}, which fuses SPIN features with DOA cues through subband-wise FiLM modulation. CORE achieves the best scores on all metrics,
improving PESQ, ESTOI, and SI-SDR~\cite{rix2001pesq,jensen2016estoi,leroux2019sisdr} over Concat
(ID~4) by 0.30, 5.22 percentage points, and 1.61~dB, respectively, with marginal overhead. It also outperforms
cross-attention with fewer MACs and parameters, validating the effectiveness
and efficiency of TF-GCA in modeling fine-grained DOA--TF interactions. Concat (large) remains inferior despite using about $3.5\times$ as many MACs
and parameters as CORE, showing that effective fusion matters more than
capacity.

\subsection{Ablation Studies and Analysis of CORE}

\input{Figures/fig4_layout}

IDs~10--14 in Table~\ref{tab:doa_conditioning} evaluate the main CE components.
Removing the spatial and spectral branches reduces SI-SDR by 0.63 and
2.35~dB, respectively; nevertheless, the latter variant still achieves
8.21~dB, suggesting that the remaining IPD pathway provides useful
directional cues. Replacing TF-GCA with direct feature concatenation while retaining LGM reduces
SI-SDR by 0.28~dB, whereas removing LGM causes a 0.45~dB drop. These results
indicate complementary contributions from bin-wise modulation and local
temporal mixing. LGM outperforms SwiGLU in
PESQ, ESTOI, and all
DNSMOS~\cite{reddy2022dnsmosp835,reddy2021dnsmos} scores, supporting the
benefit of local temporal context.

To analyze the representation learned by CE, we froze its weights after CORE training and optimized only a $1\times1$ projection for complex-mask estimation. Figure~\ref{fig:core_extraction} visualizes the extraction process:
TF-GCA forms a structured latent TF attention map during the initial direction-aware stage of DSE, CE produces a coarse
target mask, and FE recovers finer target-speech details, demonstrating the roles of coarse extraction and refinement.

\subsection{Generalization Across Fine Enhancers}

\input{Tables/table2}

Table~\ref{tab:fine_enhancers} evaluates CORE with four FEs on \textbf{Easy} and \textbf{Hard}, including state-of-the-art (SOTA) FEs under more challenging 3D multi-speaker conditions on \textbf{Hard}. Without CE, the spectral, IPD, and DOA features are
concatenated and fed directly into each FE. Adding CE consistently improves every FE on all quality
metrics with marginal MAC and parameter overhead, demonstrating the cross-backbone applicability of
CORE. Audio samples are available at \url{https://muefy.github.io/CORE-DSE-Audio-Demo/}.

\section{Conclusion}

In this paper, we propose CORE-DSE to address the granularity mismatch between global DOA cues and
dense TF acoustic representations through direction-aware coarse extraction and target-speech
refinement. CE combines TF-GCA for fine-grained DOA–TF modeling and LGM for local temporal context in spectral and spatial representations, while FE suppresses residual interference and
recovers target speech. Experiments show that CORE-DSE outperforms physics-informed and generic DOA
conditioning baselines with marginal computational overhead. Effective directional
fusion yields greater gains than increasing model capacity, while improvements across FE
backbones support the effectiveness and generality of CORE.

\section{Compliance with Ethical Standards}

No ethical approval was required for this numerical simulation study. The authors have no competing
interests to disclose.

\bibliographystyle{IEEEbib}
\bibliography{strings,refs}

\end{document}

%% file: Figures/fig1_layout.tex
\begin{figure}[t]
\centering
\includegraphics[width=0.65\columnwidth]{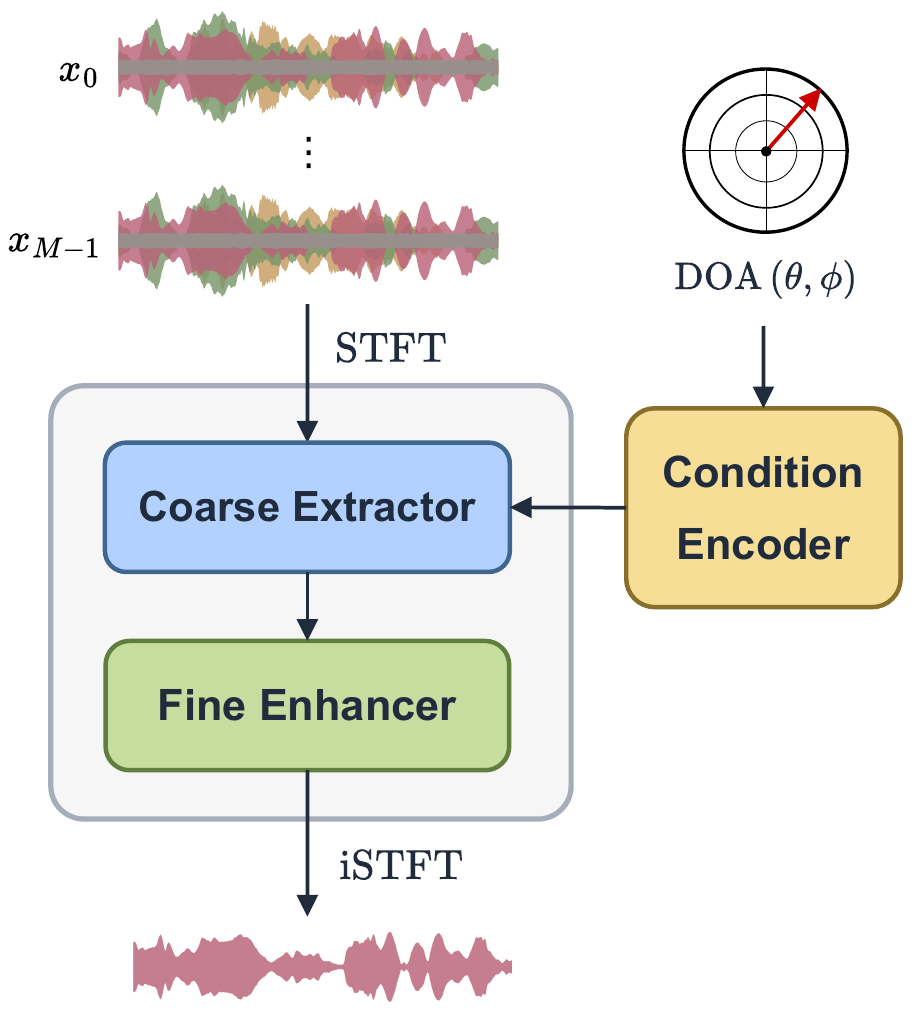}
\caption{Overview of the CORE-DSE framework.}
\label{fig:core_dse_overview}
\end{figure}

%% file: Figures/fig2_layout.tex
\begin{figure}[t]
\centering
\includegraphics[width=0.9\columnwidth]{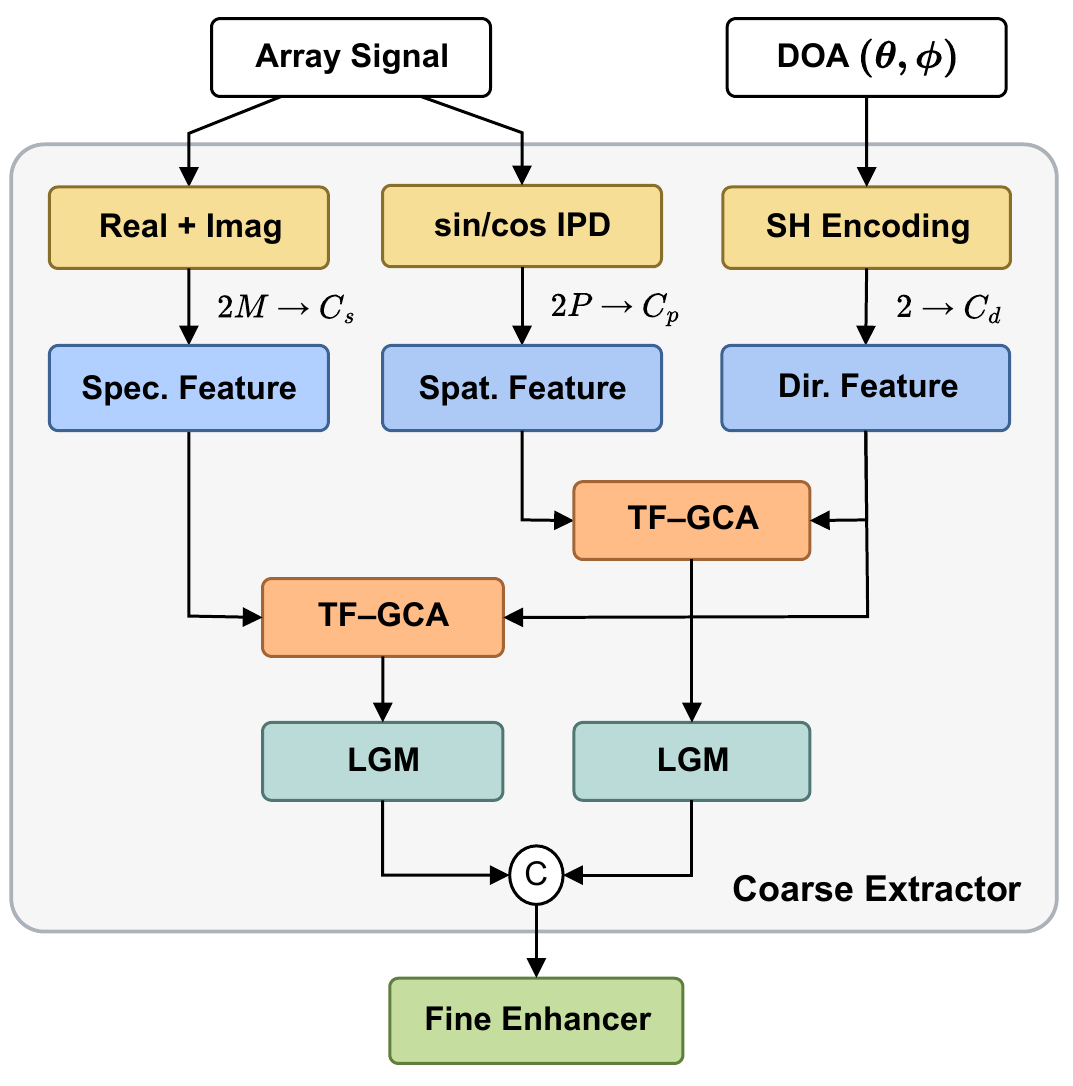}
\caption{Architecture of the Coarse Extractor.}
\label{fig:coarse_extractor}
\end{figure}

%% file: Figures/fig3_latout.tex
\begin{figure*}[t]
\centering
\subfloat[TF-GCA\label{fig:tf_gca}]{\includegraphics[height=0.26\textwidth]{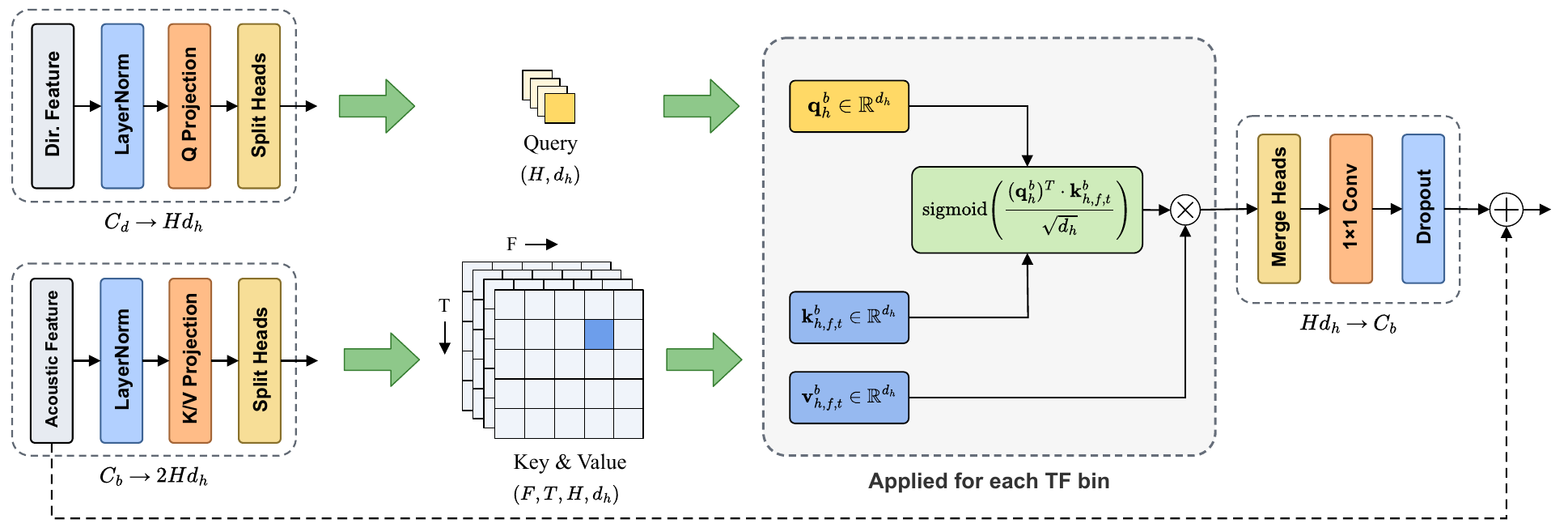}}%
\hfill
\subfloat[LGM\label{fig:lgm}]{\includegraphics[height=0.26\textwidth]{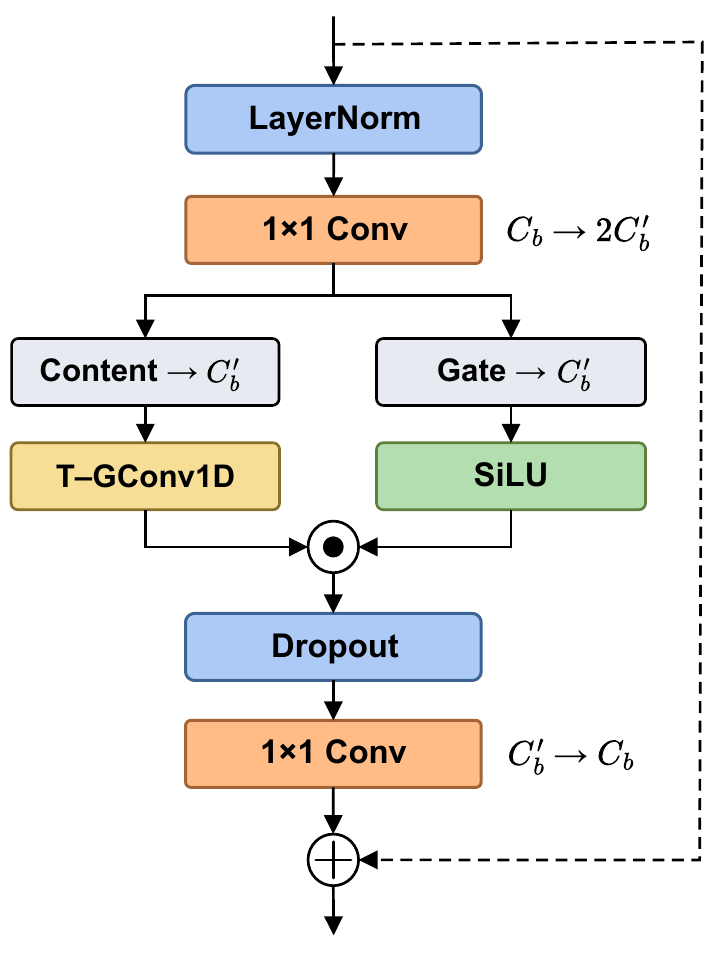}}
\caption{Architectures of the proposed modules: (a) Time-Frequency Gated Cross-Attention; (b) Local Gated Mixer.}
\label{fig:tf_gca_lgm}
\end{figure*}

%% file: Tables/table1.tex
\begin{table*}[t]
\centering
\caption{Comparison of DOA conditioning strategies and ablations on the \textbf{Easy} dataset. \textbf{Bold} values indicate the best score for each metric.}
\label{tab:doa_conditioning}
\fontsize{9}{10}\selectfont
\setlength{\tabcolsep}{3pt}
\renewcommand{\arraystretch}{0.9}
\setlength{\aboverulesep}{0.36ex}
\setlength{\belowrulesep}{0.58ex}
\begin{tabular*}{\textwidth}{@{\extracolsep{\fill}}cl@{\hspace{16pt}\extracolsep{\fill}}ccccccccc@{}}
\specialrule{\heavyrulewidth}{0pt}{0.05ex}
\multirow{2}{*}[-2pt]{ID} & \multirow{2}{*}[-2pt]{Method} & \multirow{2}{*}[-2pt]{\raisebox{-1.4pt}[14pt][0pt]{\shortstack{MACs\\[-0.5pt](G/s)}}} & \multirow{2}{*}[-2pt]{\raisebox{-1.4pt}[14pt][0pt]{\shortstack{Params\\[-0.5pt](M)}}} & \multirow{2}{*}[-2pt]{PESQ} & \multirow{2}{*}[-2pt]{\raisebox{-1.4pt}[14pt][0pt]{\shortstack{ESTOI\\[-0.5pt](\%)}}} & \multirow{2}{*}[-2pt]{\raisebox{-1.4pt}[14pt][0pt]{\shortstack{SI-SDR\\[-0.5pt](dB)}}} & \multicolumn{4}{c}{\raisebox{-2.0pt}{DNSMOS}} \\[1.4pt]
\cmidrule(lr){8-11}
\noalign{\vskip -0.4pt}
& & & & & & & OVRL & SIG & BAK & P808 \\[-0.6pt]
\specialrule{\lightrulewidth}{0.05ex}{0.50ex}
\noalign{\vskip 0.4pt}
-- & Unprocessed & -- & -- & 1.10 & 35.26 & -5.63 & 1.24 & 1.47 & 1.28 & 2.72 \\[-0.4pt]
\midrule
1 & Angle feature~\cite{chen2018location} & 23.61 & 1.45 & 2.15 & 73.06 & 8.45 & 2.52 & 2.94 & 3.54 & 3.41 \\
2 & Delay-and-sum beam~\cite{elminshawi2023beamformer} & 23.61 & 1.45 & 2.18 & 74.12 & 8.98 & 2.53 & 2.96 & 3.51 & 3.41 \\
\midrule
3 & Initial~\cite{tesch2024spatial} & 23.60 & 1.59 & 2.13 & 72.95 & 8.53 & 2.52 & 2.94 & 3.55 & 3.40 \\
4 & Concat (w/o CE) & 23.78 & 1.46 & 2.18 & 73.84 & 8.95 & 2.54 & 2.96 & 3.56 & 3.42 \\
5 & FiLM & 23.62 & 1.46 & 2.32 & 76.34 & 9.85 & 2.59 & 3.01 & 3.59 & 3.46 \\
6 & Cross-attention & 25.98 & 1.56 & 2.42 & 77.83 & 10.42 & 2.61 & 3.03 & 3.59 & 3.46 \\
7 & Concat (large) & 85.55 & 5.27 & 2.32 & 76.16 & 9.79 & 2.61 & 3.01 & 3.60 & 3.47 \\
8 & SoundCompass fusion~\cite{choi2026soundcompass} & 24.31 & 2.19 & 2.23 & 74.71 & 8.64 & 2.57 & 2.98 & 3.59 & 3.46 \\
\midrule
9 & \textbf{CORE (ours)} & 24.46 & 1.51 & \textbf{2.48} & \textbf{79.06} & \textbf{10.56} & \textbf{2.65} & \textbf{3.08} & \textbf{3.61} & \textbf{3.52} \\
10 & w/o spatial branch & 24.15 & 1.49 & 2.37 & 77.70 & 9.93 & 2.62 & 3.05 & 3.59 & 3.50 \\
11 & w/o spectral branch & 24.15 & 1.49 & 2.08 & 72.62 & 8.21 & 2.45 & 2.89 & 3.40 & 3.35 \\
12 & w/o TF-GCA & 24.34 & 1.50 & 2.40 & 77.64 & 10.28 & 2.63 & 3.05 & 3.60 & 3.50 \\
13 & w/ SwiGLU & 24.41 & 1.51 & 2.45 & 78.59 & 10.52 & 2.62 & 3.05 & 3.57 & 3.48 \\
14 & w/o LGM & 24.01 & 1.49 & 2.38 & 77.37 & 10.11 & 2.61 & 3.03 & 3.59 & 3.47 \\
\bottomrule
\end{tabular*}
\end{table*}

%% file: Figures/fig4_layout.tex
\begin{figure}[t]
\centering
\makebox[\columnwidth][c]{%
  \hspace*{-0.04\columnwidth}%
  \includegraphics[width=0.93\columnwidth]{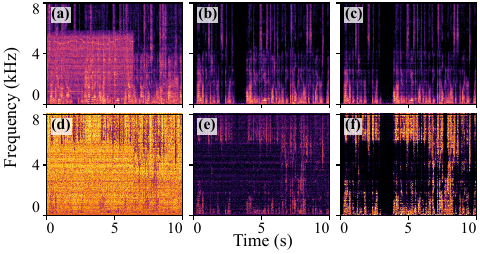}%
  \hspace*{0.04\columnwidth}%
}
\caption{Visualization of the CORE extraction process: (a) unprocessed speech; (b) clean speech; (c) CORE output speech; (d) TF-GCA map ($h=1$); (e) CE-only output mask; (f) CORE output mask.}
\label{fig:core_extraction}
\end{figure}

%% file: Tables/table2.tex
\begin{table}[t]
\centering
\caption{Performance comparison of different FEs with/without CE.}
\label{tab:fine_enhancers}
\fontsize{9}{10}\selectfont
\setlength{\tabcolsep}{2pt}
\renewcommand{\arraystretch}{0.8}
\setlength{\aboverulesep}{0.36ex}
\setlength{\belowrulesep}{0.58ex}
\begin{tabular*}{\columnwidth}{@{\extracolsep{\fill}}l@{\hspace{7pt}\extracolsep{0pt}}c@{\extracolsep{\fill}\hspace{4pt}}ccccc@{}}
\toprule
\multirow{2}{*}{FE} & \multirow{2}{*}{CE} & MACs & Params & \multirow{2}{*}{PESQ} & ESTOI & SI-SDR \\
& & (G/s) & (M) & & (\%) & (dB) \\
\midrule
\multicolumn{7}{@{}l}{\textbf{2D-Spatialized Dataset (Easy)}} \\
\midrule
Unprocessed & -- & -- & -- & 1.10 & 35.26 & -5.63 \\
\midrule
\multirow{2}{*}{EaBNet~\cite{li2022eabnet}} & $\times$ & 5.35 & 3.35 & 2.18 & 74.36 & 8.55 \\
& $\checkmark$ & 5.48 & 3.36 & \textbf{2.42} & \textbf{77.66} & \textbf{9.74} \\
\midrule
\multirow{2}{*}{FT-JNF~\cite{tesch2022ftjnf}} & $\times$ & 23.78 & 1.46 & 2.18 & 73.84 & 8.95 \\
& $\checkmark$ & 24.46 & 1.51 & \textbf{2.48} & \textbf{79.06} & \textbf{10.56} \\
\midrule
\multicolumn{7}{@{}l}{\textbf{3D-Spatialized Dataset (Hard)}} \\
\midrule
Unprocessed & -- & -- & -- & 1.15 & 23.55 & -9.83 \\
\midrule
\multirow{2}{*}{TF-GridNet~\cite{wang2023tfgridnet}} & $\times$ & 21.74 & 1.30 & 2.09 & 70.21 & 7.82 \\
& $\checkmark$ & 22.42 & 1.35 & \textbf{2.20} & \textbf{72.34} & \textbf{8.38} \\
\midrule
\multirow{2}{*}{SpatialNet~\cite{quan2024spatialnet}} & $\times$ & 25.35 & 1.74 & 2.72 & 81.00 & 11.67 \\
& $\checkmark$ & 26.02 & 1.79 & \textbf{2.86} & \textbf{83.17} & \textbf{12.57} \\
\bottomrule
\end{tabular*}
\end{table}